# Anisotropic Rheology and Friction of Suspended Graphene

*Andrea Mescola[1], Andrea Silva[2,3], Ali Khosravi[2,3,4], Andrea Vanossi[2,3], Erio Tosatti[2,3,4], Sergio Valeri[1,5] and Guido Paolicelli[1*]*

[1] CNR-NANO, Consiglio Nazionale delle Ricerche - Istituto Nanoscienze, Via Campi 213 41125 Modena, Italy.

[2] CNR-IOM, Consiglio Nazionale delle Ricerche - Istituto Officina dei Materiali, c/o SISSA Via Bonomea 265, 34136 Trieste, Italy.

[3] International School for Advanced Studies (SISSA), Via Bonomea 265, 34136 Trieste, Italy.

[4] The Abdus Salam International Centre for Theoretical Physics (ICTP), Strada Costiera 11, 34151 Trieste, Italy

[5] Department of Physics, Informatics and Mathematics, University of Modena and Reggio Emilia, Via Campi 213 41125 Modena, Italy.

[*] *corresponding author: guido.paolicelli@nano.cnr.it*

**ABSTRACT**

Graphene is a powerful membrane prototype for both applications and fundamental research. Rheological phenomena including indentation, twisting, and wrinkling in deposited and suspended graphene are actively investigated to unravel the mechanical laws at the nanoscale. Most studies focused on isotropic set-ups, while realistic graphene membranes are often subject to strongly anisotropic constraints, with important consequences for the rheology, strain, indentation, and friction in engineering conditions.

Graphene in particular is recognized as the thinnest solid lubricant material and a large amount of work has been dedicated to understanding the fundamentals mechanisms of this effect and to unravel parameters relevant to its technological development. Here we experimentally show how graphene's frictional response to an external indenter is severely altered by conditions of anisotropic suspension, specifically when graphene is clamped across a long and narrow groove. Results show that the friction coefficient is significant when the tip is sliding parallel to the groove while becoming ultra-low in the orthogonal direction. While the experimental data suggest that – rather unexpectedly – pre-strain of the graphene sheet as a result of clamping is negligible, the key to understanding the underlying mechanism is provided by simulations. The paramount mechanism is provided by the extra anisotropic strain induced from indentation under anisotropic constraints, which in turn produces an anisotropic stiffening of the graphene. While the focus of this work is on graphene, we believe our experimental protocol and the physical mechanism uncovered by our model can be applicable to other 2D membrane-like materials.

**Introduction**

The rheological and frictional behavior of pristine graphene has attracted much fundamental and technological interest during the last decade. From the mesoscale down to the atomic level, a great research effort is underway to unveil the physical mechanisms underpinning the indentation, twisting, wrinkling and crumbling phenomena [1–8] of this extraordinary membrane. The exceptional hardness (Young's modulus around 1 TPa and intrinsic in-plane strength of 130 GPa [5]), the extreme ability to elastically sustain tensile strain up to at least 20% [9], and its large out-of-plane membrane-like flexibility make graphene the forefront prototype material for the design of innovative systems and structures with controlled intrinsic properties [10]. Graphene has been integrated into several hi-tech electronic devices such as organic light-emitting diodes (OLED) [11,12], strain sensors [13–17], wearable devices [18–20] and micro- and nanoelectromechanical systems (MEMS and NEMS) [21]. Understanding and exploiting strain-induced effects in 2D layered materials in general is nowadays a fast-growing topic in both fundamental and applied research, given that many of its intrinsic properties can be tuned by mechanical deformation. While the tunability of transport and optical properties under strain has been extensively addressed [22–26], the microscopic tribological behavior as a function of the applied strain and boundary conditions has received far less attention [27–29]. Graphene tends to conform to the morphology of the supporting surface when deposited over a hosting substrate and the tribological response of the layered coating significantly depends on the layer-substrate interaction [30–32]. At the same time, different experimental realizations of supported graphene have highlighted the importance of 2D layer strain on the overall tribological and electronic behavior [33–35]. To disentangle these effects and shed light on their origin, suspended graphene represents an ideal system. Recently, free-standing graphene has been investigated by S. Zhang *et al*. [27], who found a reduction of friction with increasing strain in a suspended and isotropically strained graphene sheet with Atomic Force Microscopy (AFM) measurements. Complementary atomistic simulations showed how this tensile pre-strain reduces the membrane flexibility thus decreasing the impact of the tip, hence reducing the friction.

Here we go beyond the isotropic setup in Ref. [27] and study the  friction behavior of graphene suspended and fixed at the edges of a long narrow groove. We observe with remarkable reproducibility that the frictional dissipation of the graphene membrane measured by our sliding AFM tip turns anisotropic, with a differential friction coefficient (COF) typically three times higher parallel to the groove axis compared to orthogonal axis. This occurs with negligible pre-strain in our system, as confirmed by Raman measurements. Molecular-dynamics (MD) simulations reveal that the sliding friction anisotropy is ruled by the interplay between the tip indenting action and the boundary conditions. The strain-free, asymmetric suspension condition is responsible for the anisotropic indentation pattern shape and rheology of the membrane, quantitatively explaining the observations. Our results rationalize the often-overlooked aspect of non-isotropic constraints in nanoscale systems, with important implications for realistic engineering setups.

**Sample characterization**

The suspended graphene was obtained by depositing a commercial single layer CVD graphene on a  standard silicon dioxide-based calibration grating array.

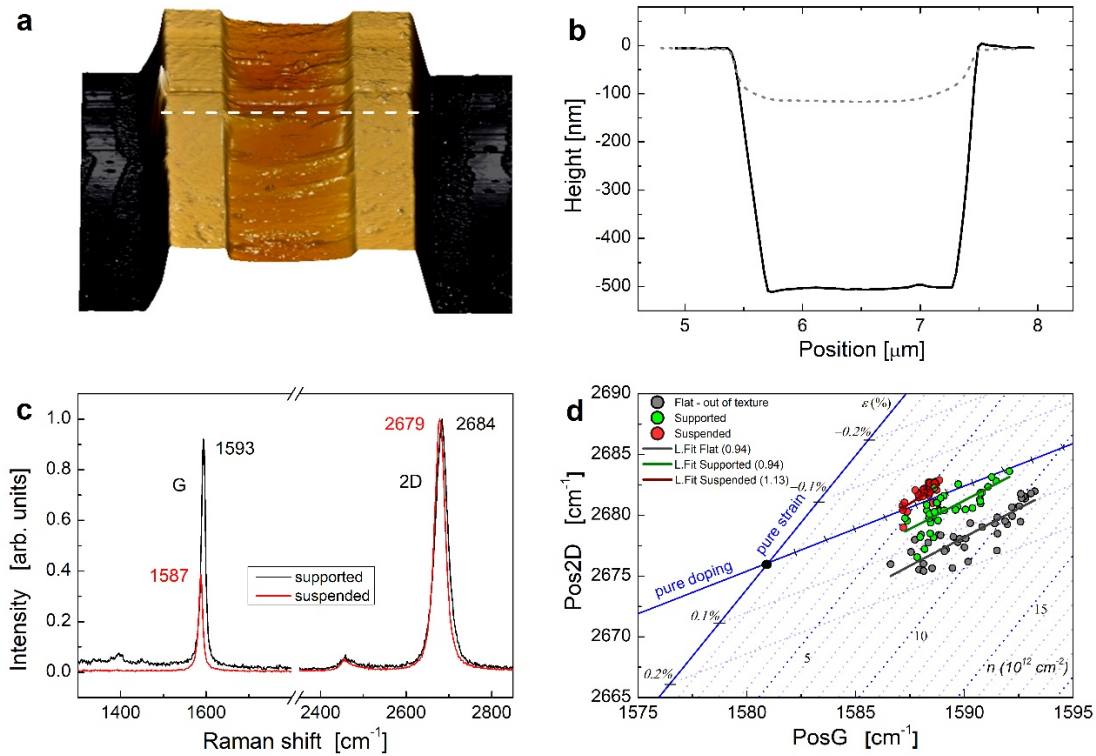

**Figure 1**. AFM imaging and Raman characterization of single layer CVD graphene anisotropically suspended. (a) 3D topography image and (b) relative line profiles of CVD graphene deposited on a single groove. The white dashed line in panel (a) indicates the line profile in panel (b). The equilibrium configuration between graphene and underlying substrate indicates a complete conformation and clamping on the flat crests enclosing the groove and a total suspension at the center of the groove. (c) Raman spectra comparison of supported graphene (black) and suspended graphene (red). Note the typical redshift of G and 2D peak positions on suspended graphene. (d) Correlation plot of Pos2D versus PosG peaks showing data from supported graphene (gray) and data within the patterned area which include supported graphene on the crest and suspended graphene (green and red, respectively). Lines represent the linear fit from supported graphene (gray line, slope 0.94), graphene on crest (green line, slope 0.94), and from suspended graphene (red line, slope 1.13). The neutrality point indicated by the black dot at position (1581.6 cm$^{-1}$, 2676.9 cm$^{-1}$) in the Pos2D versus PosG correlation plot, and the additional axes to quantify strain and doping are taken from Ref. [36].

A schematic representation of the sample construction is shown in Supporting Information [37]. Briefly, a commercial single layer CVD graphene was deposited over the calibration grating by wet transfer process. The calibration grating comprises a central patterned area and a surrounding flat region. The pattern consists of long, parallel and equally spaced grooves (crest-to-crest distance: 3 µm, valley width 2 µm, step height: 500 nm). After transfer, the external flat region is fully covered by single layer graphene that we refer to as supported graphene in the following. On the contrary, the sample surface over patterned area displayed regions where the graphene adhered to the substrate and regions where it hung fully suspended via clamping on the top of the crests (Fig. 1a). AFM 3D-topography and corresponding line profiles in Fig. 1a, b, reveal an excellent conformation of graphene sheet over the crests and the complete suspension in the region between them. In particular, the AFM profiles before and after deposition (solid and dashed lines Fig. 1b, respectively) indicate that graphene tends to mechanically relax to a configuration where a small fraction adheres to the side walls of the groove.

The graphene membrane was then characterized by Raman spectroscopy, a well-established method used to quantify the purity, thickness and strain of graphene films [36,38–40]. Figure 1c presents the typical Raman spectra obtained from analyzing both the supported graphene (black line) and suspended graphene (red line) of the deposited graphitic film. The shape of the two prominent G and 2D peaks (both symmetric Lorentzian line shape, and width $W_G = 12$ cm$^{-1}$ $W_{2D}$ = 33 cm$^{-1}$ respectively) indicates a high-quality, single-layer suspended graphene film [38–41]. The absence of the disorder-related D peak near 1350 cm$^{-1}$ on the suspended part and the weak intensity on the supported one indicate that CVD graphene film is composed of large, defect-free single-crystal domains. The downward shift of both peaks positions in the suspended regions with respect to the supported graphene is evident and indicative of the different extent of both deformation and doping effects [38] in the two regions.

To disentangle doping and strain effects, we performed extensive Raman maps on graphene on the patterned area and graphene supported on the surrounding flat silicon region. Each map comprised about 80 points where we measured and fit Raman spectra to obtain G and 2D peak positions (PosG and Pos2D in the following). We computed the Pos2D/PosG correlation shown in Fig. 1d following the protocol developed by Lee *et al.* [36]. PosG and Pos2D are linearly correlated even within an ideal graphene sheet because of the ubiquitous presence of doping and strain effects. The slope of the linear distribution is indicative of the relative importance of either doping or strain effects. The solid blue lines in Fig. 1d represents the linear pure strain axis with slope 2.2 and the pure doping axis with slope 0.7, respectively [36]. The two axes cross at the neutrality point marked by a black dot at position (1581.6 cm$^{-1}$, 2676.9 cm$^{-1}$) [36]. While on the doping line only positive values are physically meaningful, the convention adopted for strain evaluation is to indicate as a negative value the compressive strain that moves the peak towards higher frequencies and as a positive value the tensile strain inducing shift in the opposite direction [42]. The gray dot and the related gray line represent data from supported graphene while green and red data come from graphene over the patterned area. Data from supported graphene (gray symbols) indicate only a very slight tensile strain as well as a hole-doping effect, consistent with the literature for supported CVD graphene [39]. Raman spectra acquired from graphene within the pattern split into two sets with different slopes (green and red symbols). The green dots show a trend compatible with the reference signal (supported graphene) and, hence, they are attributed to the crest region, where graphene is supported. Indeed, the fitted slope (green and gray lines in Fig. 1d) is 0.94 in both cases, a value consistent with the literature [39]. By contrast, the red data are representative of suspended graphene: they group close to the doping axis so that the extrapolation towards strain axis indicates a negligible pre-strain. Indeed the linear fit yields a slope of 1.13 consistent with the value of 1.09 reported by Gajewski et al. [39] for suspended graphene. Hence, our sample presents no pre-strain, probably because the portion of membrane adhering to the side wall of the groove is small and its effect negligible.

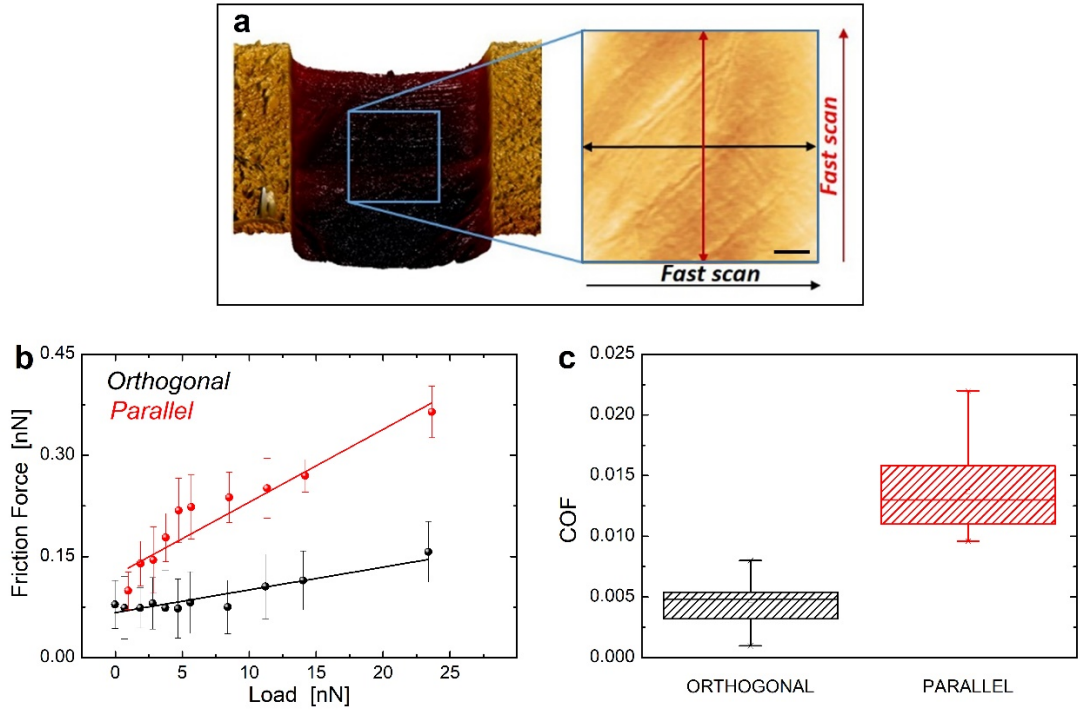

**Figure 2.** Direction dependent friction response of suspended CVD graphene. (a) 3D topography reconstruction of freestanding CVD graphene over single groove and 2D zoom of the region analyzed by FFM (blue square: 1×1 μm²); the two scan directions are represented with double-headed arrows (black for the orthogonal and red for the parallel, respectively). Scale bar corresponds to 150 nm. (b) Friction force as a function of load applied to freestanding CVD graphene with groove axis oriented orthogonal (black) and parallel (red) to the fast scan direction; circles represent experimental data with their error bars and continuous lines are the respective linear fit. (c) Boxplots of COF for orthogonal (black) and parallel (red) scans; counting corresponds to 12 different regions for orthogonal scans and to 19 for parallel scans, respectively.

**Results and Discussion**

Following the structural results of the Raman spectroscopy analysis, the membrane rheology was proved by Friction Force Microscopy (FFM) (Fig. 2). FFM scans in two opposite directions were performed on a one square micron area, corresponding to the central region suspended between two crests (blue square in Fig. 2a). By tilting the sample 90 degrees, the same area was first scanned in the direction orthogonal to the groove axis and then, in the one parallel to the groove axis, as indicated in the right side of Fig. 2a by black and red arrows, respectively. The same area was analyzed in both directions with positioning accuracy of the order of 100 nm ( [43,44] and details in Supporting Information [37]).

Care has been taken to analyse regions free from pre-existing extended wrinkles, which are clearly identifiable due to the strong contrast they produce on the lateral force signal [45]. On the contrary, ripple effects on graphene either of an intrinsic type or induced by tip or point defect [46] cannot be resolved since the induced elastic deformations during FFM measurements are larger with respect to ripple corrugation.

The measured frictional dissipation F was found to depend strongly on the scanning direction: the friction force recorded during scans parallel to the groove was typically three times that the orthogonal, at the same load P (see Fig. 2b and *Fig. S2a*). The orthogonal *vs* parallel friction increase was then measured at different loads, from about 25nN down to the negative pull-off force. Figure 2b reports the typical friction *vs* load behavior observed in several different spots on the membrane. The differential friction coefficient dF/dP (COF) was evaluated for each friction *vs* load curve and used as a representative parameter of the membrane anisotropic response. Fig. 2c summarizes the statistics of our multi-spot analysis, revealing a remarkably consistent three-fold increase from orthogonal to parallel scanning direction.

In particular we notice that along the orthogonal direction the extremely low COF values are in agreement with those measured by Z. Deng et al. on single, bi and tri layer graphene suspended on micrometre size circular hole [47] and confirmed recently by S. Zhan [27]. These values approach that of thick graphite [48] suggesting the absence of important elastic deformation effects. Comparison with graphene supported systems is non-trivial since the specific tribological behaviour depends on the substrate on which the graphene is deposited [27,49]. In general, the higher the adhesion towards the substrate the lower the out of plane deformation effect and the lower the friction coefficient [50,51]. Our results, summarised on fig. 2 panel c, reveal that along the orthogonal direction the membrane behaves like a graphene layer with minor load induced out of plane deformations while along parallel direction load dependent deformation effects seems becoming increasingly important.

The FFM scans were performed over hundreds of nanometers, providing the mesoscopic tribological response arising from a nanoscale contact. Over this length scale the atomistic stick-slip events cannot be detected by our equipment. Nevertheless, the atomistic nature of the dissipation can be indirectly assessed from the velocity dependency of the COF. We found that halving and doubling the sliding velocity yielded the same COF (Supporting Information [37]). This velocity-independence of friction in both directions is the fingerprint of the stick-slip regime, whereas a viscous dissipation would yield a linear dependence [52].

In order to understand the atomistic mechanism underlying the microscopic experimental measures, we consider the minimalistic model sketched in Fig 3a, which retains the experimental features assumed to be at the origin of the observed behavior. The stick-slip regime indicated by the experiments suggests that the fundamental dissipation mechanism should be captured by the Prandtl-Tomlinson (PT) model [53,54], allowing for the tip to be reduced to a point-like object (blue dot in Fig. 3a) sliding over a corrugated energy landscape. The cantilever was modeled as a mass moving at constant velocity $v_{drag}$ (red square in Fig 3a), dragging the tip via a spring of constant K(see Supporting Information [37] and a benchmark of the model parameters at Refs. [27,55,56].

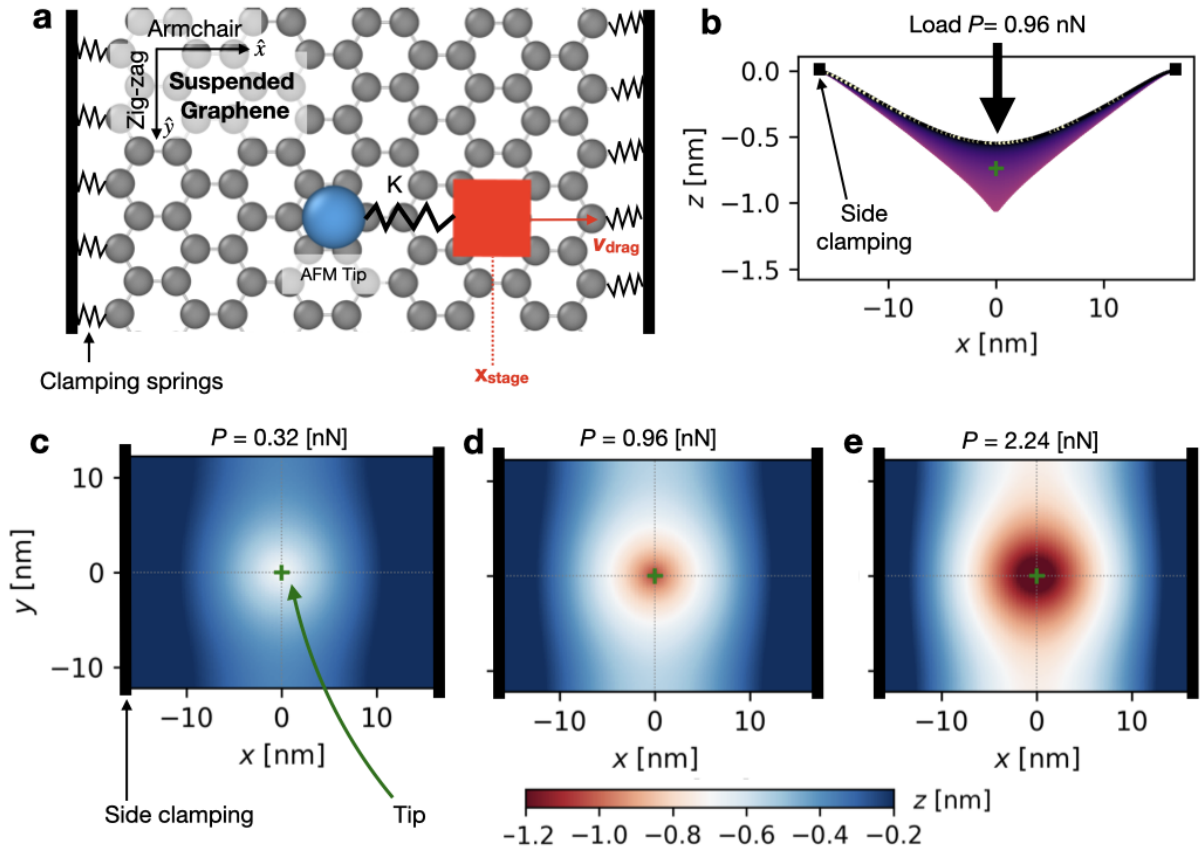

**Figure 3.** (a) Model setup. The free-standing graphene membrane is clamped along the edges in the $y$ direction at fixed $x$ coordinate, consistently with experimental orientation in Fig. 1a. The clamping is realized by attaching a stiff spring ($Q$=1602 N/m) to each edge carbon atom in a 0.25 nm region. As reported in the top left corner, the zig-zag edge is clamped while the armchair one is free (see SI for results with the opposite orientation, namely clamped armchair and free zig-zag). The Prandtl-Tomlinson tip (blue sphere) indents the membrane under a constant vertical load (along $z$). Mimicking a minimal AFM setup, the tip is attached to a moving stage (red cube), representing the massive cantilever, translating at constant velocity $v_{drag}$. Side (b) and top (c, d, e) views of the indentation profile. The purple region in (b) reports the position projected on $x$ and the black squares indicate the position of the clamps. Note that the axes in (b) are not in scale, i.e. the deformation of the membrane is accentuated for visual aid. (c, d, e) Top-view of the indentation profile for increasing loads, as reported at the top of each plot: the tip (green cross) indents the center of the membrane and the system is relaxed (see SI for the relaxation protocol). The color scale in (c-e) reports the vertical deformation of the membrane, as indicated by the color bar at the bottom. The black bar at the sides indicates the position of the clamps, as in (a). The gray dashed lines in (c-e) mark the reference x=0, y=0, as guide to the eye. The height of each C atom of the membrane has been linearly interpolated on a finer grid for clearer visualization.

Construction of the substrate in the PT model (i.e. the 2D membrane on which the tip slides) was key in this model, as it needed to capture the deformable nature of the suspended graphene sheet while at the same time preserving the atomistic nature of the contact – a smooth membrane cannot yield a stick-slip dynamics and a discrete graphene membrane without the mesoscopic clamping asymmetry cannot deform realistically under the tip. To achieve a reasonable tradeoff between these two opposite requirements, we modeled the experimental membrane as a classical graphene sheet where the AIREBO [56] potential describes intra-layer interactions and a Lennard-Jones (LJ) potential [27] interactions between the tip and graphene. To reproduce the asymmetric geometry, the membrane was finite in the $x$ direction (orthogonal), and periodic in the $y$ direction (parallel). To mimic the adhesion to the groove's crests, the membrane edges in the $x$ direction were clamped with springs of constant $Q$=1602 N/m to the equilibrium position, marked by black straight line in Fig. 3a. Hence, simulated membrane was suspended and the clamping springs provided the restoring force opposing the one exerted by the tip upon loading and sliding. In Fig. 3a the clamped side is the zig-zag edge while the periodic side is the armchair edge. Effects of finite size, edge orientation and full open boundary conditions are discussed in the SI. The simulations were performed at zero temperature to obtain a clear signal and because the stick-slips dynamics is regulated by the energy barrier, while temperature introduces thermo-lubricity without qualitatively modifying the dynamics [53,54]. As the tip in experiments moved orders of magnitude slower than the relaxation time of the atomistic motion, a viscous damping was applied to all the atoms. The crucial property we rely upon is the near independence of stick-slip friction upon velocity [52].

An external load $P$ was applied to the model tip (as sketched in Fig. 3b), which indents the membrane, as shown in Fig. 3c-e for increasing loads. The indentation shape follows the asymmetric geometry. The deformation is radial under the tip (green cross in Fig. 3c-e) but this symmetry is lost far away from it, in particular nearing the edges: the membrane forms an elongated "valley" in the parallel direction (white regions in Fig. 3c-e) and remains almost flat near the clamps (blue regions in Fig. 3c-e). This characteristic indentation shape becomes more evident as the load increases, as shown in going from Fig. 3c to Fig. 3e.

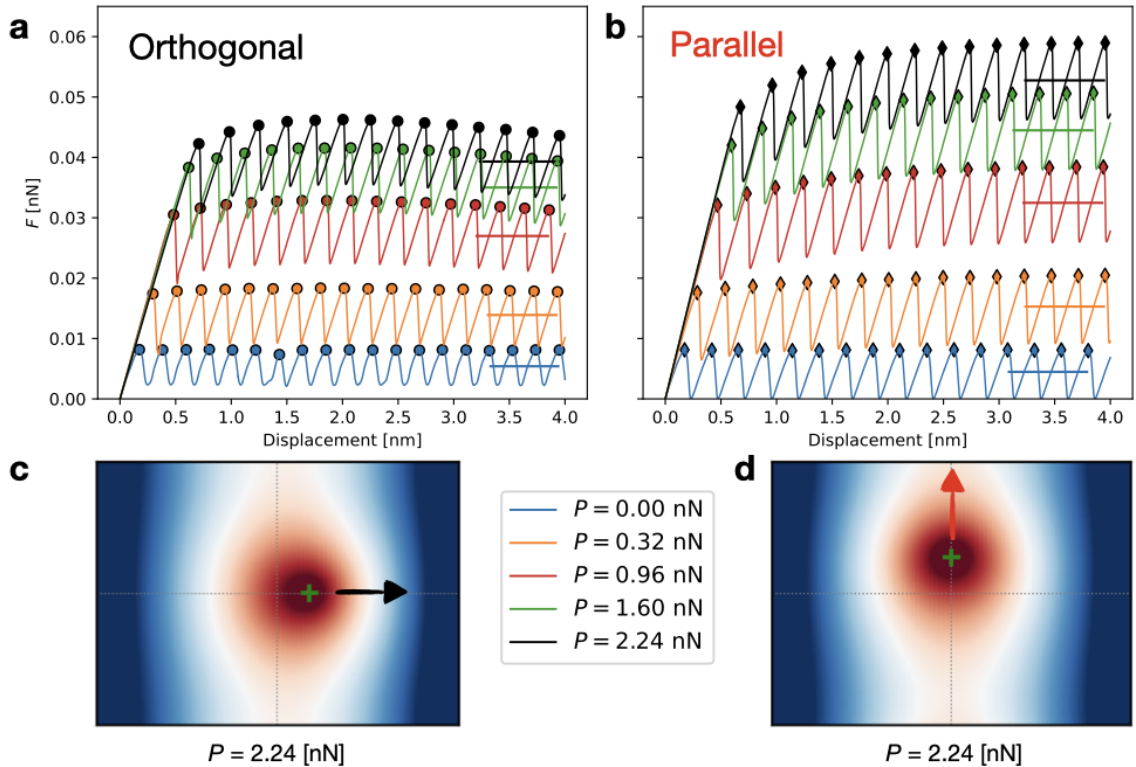

**Figure 4**. Simulated frictional response of the membrane perpendicular (a) and parallel (b) to the clamps. The y axis reports the force on the cantilever while x reports the cantilever position. Each colored line corresponds to a load, as reported in the legend in the bottom right. The friction traces are normalized by the elastic charging of the membrane as reported in SI. Color-matching horizontal lines show the average friction force at each load after the initial transient. Note that the average excludes the end of the simulation to limit the influence of the clamps in (a) and of the free-edge in (b). Panels (c), (d) show the tip sliding direction and the evolution from the indentation profile in Fig. 3e toward the end of the trajectory; the color scale is the same as Fig. 3. The dashed horizontal and vertical gray lines indicate x=0 and y=0 as a guide to the eye.

For each indentation load, the tip can slide in the orthogonal and parallel directions. The sliding was performed for a load from 0 to 2.24 nN. Note that the loads applied to a point-like tip do not relate directly to the experimental one. One can estimate the relative pressure and membrane *vs* tip size ratio between MD and experiments. Taking the equilibrium tip-membrane distance according to the LJ potential $R$=0.35 nm as tip radius in MD and assuming an Hertzian contact area (see Supporting Information [37] and Refs. [57,58]), at the maximum load $P$=2.2 nN the pressure was 12.8 GPa, much higher than experiments where the maximum estimated pressure was 28 MPa at load 20 nN. At the same time, in the experiments the load was applied on an extended tip of radius $R$=15 nm, while the real contact area of the single asperity undergoing stick-slip is certainly smaller (and thus pressure is higher) but difficult to estimate [59]. A more meaningful quantity to compare between simulations and experiments is the ratio between the tip radius R and the membrane clamped length $L$. This ratio is $R/L$=0.011 in simulations and $R/L$=0.01 in experiments, with a groove distance $L$=1.5 $\mu$m. The two values are comparatively close, suggesting that while the nominal pressure in the MD simulations is higher than the experimental one, the dimensional ratio is correctly described by the model. This geometrical element is crucial to address the anisotropy observed in experiments because the size relationship between the indenter and the intended membrane sets the shape of the indentation pattern and the strain distribution.

Force traces along the orthogonal and parallel directions for selected loads are reported in Fig. 4a, b, respectively. The atomistic stick-slip behavior is evident at all loads, owing to the damped, zero-temperature dynamics of the MD simulations. As the cantilever started to translate, the tip climbed the corrugation energy barrier, see the initial increase of all lines in Fig. 4a, b. Once the cantilever restoring force overpowered the barrier, the system depinned, causing the first drop in all lines in Fig. 4a, b. This is the onset of static friction. Afterwards, the tip moved by single-lattice slips, yielding the characteristic saw-tooth force profile of stick-slip dynamics shown in Fig. 4a, b. As shown in details on Supplementary Information [37] , the membrane deformation instantaneously follows the tip motion: there is an adiabatic separation between slow motion of the AFM tip and the fast relaxation of the membrane, as in experiments.

After an initial transient, the kinetic friction was calculated as the average sliding force shown in Fig. 4a, b as horizontal lines. The protocol to compute kinetic friction in this clamped system is described in detail in the Supplementary Information [37].

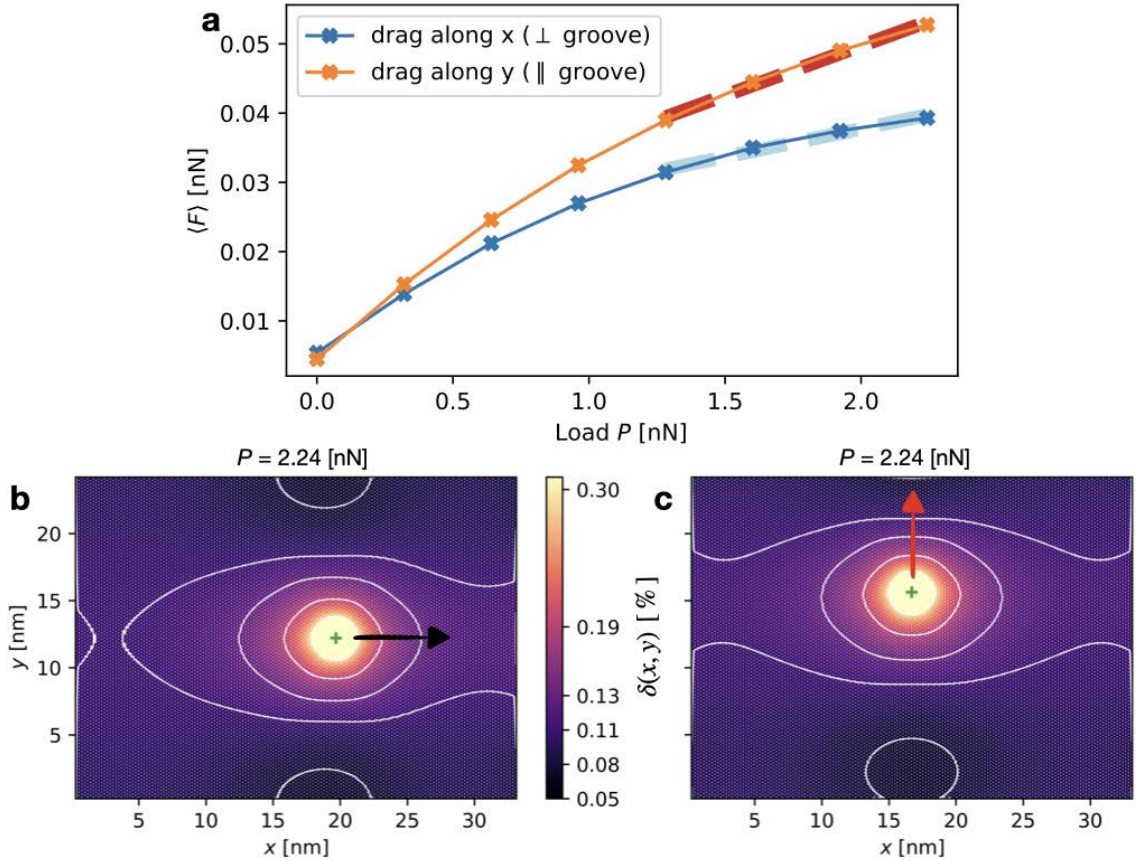

**Figure 5**. (a) Average friction versus load for the direction perpendicular (blue curve) and parallel (orange) to the clamp. The COF perpendicular and parallel to the clamps is defined as the slope of the light-blue and red dashed lines, respectively. (b), (c) Overall strain of each carbon atom as bond length difference averaged over its nearest neighbors. (b) reports the final snapshot of orthogonal dragging, (c) for parallel dragging; both snapshots refer to load P= 2.24 nN (black curves in Fig. 4a, b and last point in Fig 5a). The green cross marks the tip position. Darker regions mark smaller strain, lighter regions higher strain. White lines mark the isoline at the values reported on the color bar.

From Fig. 4a, b one can already see that the friction in the orthogonal direction is smaller than in the parallel one, as in experimental curves in Fig. 2b. The force-load curves for the two directions are obtained by plotting the average friction force, horizontal lines in Fig. 4a, b, against the applied load $P$. Figure 5a reports the average friction *vs* load for the orthogonal (blue line) and parallel (orange line). For small loads ($P$=0-0.5 nN), where the anisotropic mechanical constraint of the clamping is very modestly perceived, the two curves behave similarly; as the load increased however, the average force in the parallel direction kept rising, while in the orthogonal direction the growth slowed down. This follows the experimental behavior in Fig. 2b, where parallel force is higher while perpendicular is smaller. Even though the model force-load curves in Fig. 5a showed a clear non-linear trend, as reasonable for the system, the concept of differential COF helps to establish a systematic comparison with experiment. Considering only the large-load limit (dashed lines in Fig. 4a), we obtained $COF_\perp = 0.013$ and $COF_{II} = 0.023$, yielding a ratio $COF_{II}/COF_\perp = 1.7$. We conclude that this minimal model is able to reproduce fairly well the experimental finding $COF_{II} \gg COF_\perp$.

Note that the COF from MD simulations are higher than in experiments. This discrepancy can be linked to the system being an ideal crystal at zero temperature, representing an upper bound for the friction force in the real system. At the same, the agreement between trends found in experiments and in MD simulations are robust against variations of crystal orientation and boundary conditions (see Supplementary Information [37]and [60]). Hence, we believe the physics underpinning the experimental results are well described by the theoretical model, while the assumptions of the model result in a qualitative rather than quantitative agreement.

These "in-silico frictional experiments" were crucial to understand the origin of this asymmetric rheological response of suspended graphene in absence of pre-strain. Figure 5b, c reports the model strain distribution in the membrane as the bond length deviation from equilibrium, $< \delta >_{nn} = < \frac{l-l_0}{l_0} >_{nn}$ where $l_0 = 0.139$ nm is the equilibrium C-C bond length in MD and the $< \cdot >_{nn}$ indicates the average over the nearest neighbors of each carbon atom. While no pre-strain is present in the membrane, the tip indentation stretches the membrane up to $\delta \sim 2\%$ right below the tip as shown by the large yellow patch in Fig. 5b, c. This tip-induced strain is larger in the orthogonal direction, as the graphene needs to comply with both the tip pressure and the clamps pull (see the brighter horizontal strip highlighted by the white contour lines in Fig. 5b). Hence, the membrane is stiffer in the transverse direction. On the other hand, in the parallel direction the tip-induced strain is lower (see the black regions enclosed by the white contour lines in Fig. 5c). When the tip was dragged along the transverse direction, the stiff substrate was not deformed significantly by the pushing tip and the force needed to slide was small. Conversely, when the tip was driven along the parallel direction the graphene membrane deformed easily under the action of the sliding probe, resulting in a higher force opposing sliding and, thus, a higher dissipation. As the load increases, the orthogonal direction becomes stiffer, enhancing this effect and leading to the smaller coefficient of friction shown in Fig. 5a.

The computational results indicate that the anisotropic stiffening of the membrane induced by the clamping during indentation, and not an anisotropic pre-strain, which is absent in our case, is the mechanism underpinning the different frictional response in the orthogonal and parallel directions. This argument is further corroborated by the experiments. Analyzing the effective piezo elongation as a function of the applied loads during the acquisition of friction *vs* load images (see Supporting Information [37]) we can extract the tip-induced elastic deformations of graphene along both the orthogonal and parallel directions of the groove. This analysis provided an estimate of the effective elastic constant ($K_{eff}$) for the coupled graphene-cantilever system. The mechanics is that of two springs in series: one is given by the flexural constant $K_N$ of the AFM cantilever, the other represents the elastic response of the graphene membrane subject to normal load, whose constant is termed here $K_{Gr}$.

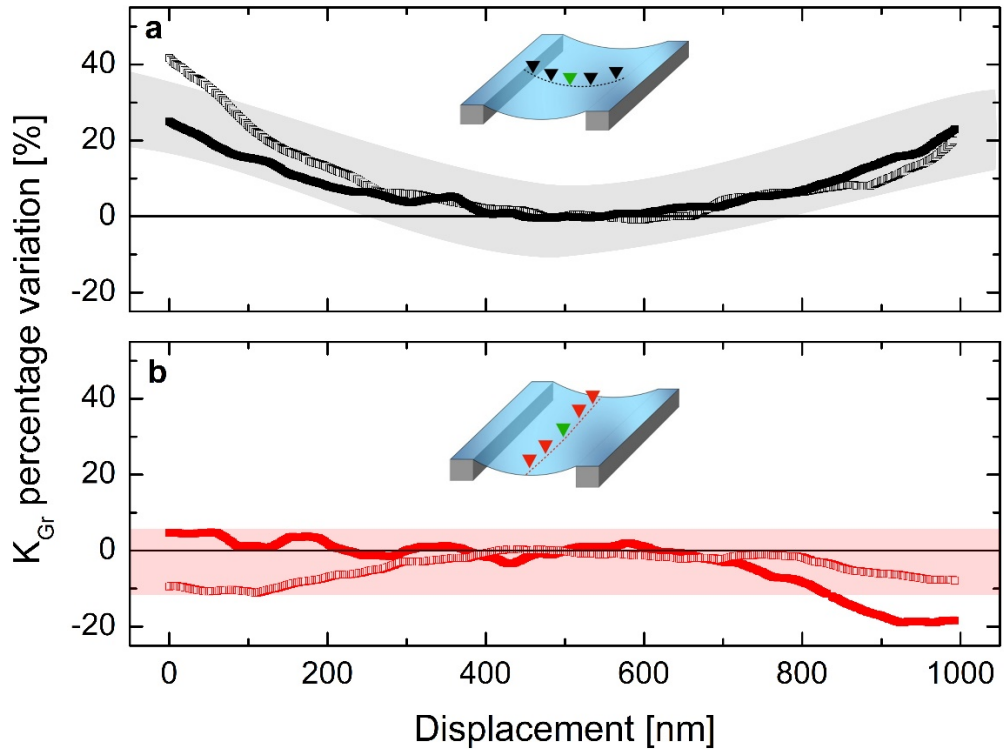

**Figure 6**. Elastic behavior of the graphene membrane along the orthogonal (black) and parallel (red) scan directions (light and dark lines correspond to different membranes). The elastic constant of the graphene membrane is evaluated as discussed on Text and on Supporting Information [37]. Data are normalized to the values evaluated at the membrane center (green triangle in the insets) and they are shown as relative percentage variation. In the orthogonal direction the relative variation of $K_{Gr}$ near the clamping regions achieves 40% while along the parallel direction it is almost always contained within 10% . The reference values at the centre are : Membrane 1 (dark line) Kgr = 0.455 ± 0.05 (N/m); Membrane 2 (light line) Kgr = 0.46 ±0.15 (N/m).

Fig. 6 reports the variation of the graphene elasticity $K_{Gr}$ as a function of the positions along the orthogonal (Fig. 6a) and parallel direction (Fig. 6b) with respect to the value at the center of the path. In the orthogonal direction, the relative variation of $K_{Gr}$ near the clamping regions achieves 40%, while along the parallel direction the variation is almost negligible (within 10% of the value at the center).

Hence, the intrinsic membrane elasticity as described by $K_{Gr}$ strongly depends on the scan direction. In particular, the orthogonal scan reveals a non-uniform response along the profile: the center region is softer while the system gradually becomes effectively stiffer approaching the clamping constraints. On the contrary, in the parallel scan direction, the response is essentially constant, as expected from the model. The behavior described by the intrinsic membrane elasticity is consistent with computational results: the induced-strain reported in Fig. 5b, c indicates anisotropic membrane stiffness along the two sliding directions. Tip sliding along the orthogonal direction (Fig 5b) moved toward higher strain regions near the clamps compared to the center: the membrane became stiffer. Conversely, when the tip slides along the parallel direction (Fig. 5c) the strain distribution just translated unchanged with the tip, yielding the same stiffness along the path.

**Conclusions**

The effect of in-plane strain on the frictional mechanics of free-standing 2D membranes has been addressed only recently. Within the effort of understanding the nano-rheology of 2D materials under strain, graphene has been investigated in different conformations, from freely suspended on a circular hole to strained as the hole gets pressurized. We found unexpected results on the asymmetric tribological response of a single layer CVD graphene freely suspended over a specially designed groove geometry. FFM measurements and friction *vs* load analysis performed at the membrane center reveal a remarkably large anisotropy with respect to the groove axis. A very low COF is measured by sliding transverse to the groove while an unexpected nearly threefold increase is obtained when moving parallel. As the pre-strain discussed in the literature is absent in our system, a different mechanism is required to explain the observed anisotropy. Our MD model, corroborated by subsequent measurements, suggests that deformation induced by indenting and sliding action of the tip is anisotropic due to the asymmetric clamping conditions. This also reflects into anisotropic friction response of the graphene membrane. Sliding orthogonally to grooves produces a larger strain, whence the membrane stiffens and the force needed to drive the tip drops. On the contrary, moving parallel to grooves the graphene membrane deforms easily and a larger force is needed to slide. This mechanism is amplified with increasing load, which results in an asymmetric dependence of friction *vs* load.

These experimental results complemented by numerical simulations demonstrate how strong adhesion (clamping) and decoupling (suspension) from the substrate can modulate the stiffness and the friction on these membrane-like systems. Strongly anisotropic geometries, such as that investigated here for the first time to the best of our knowledge, are likely to occur in realistic systems. Hence, we believe that our results and analysis will be relevant to design and tune future graphene-based systems such as nano-mechanical devices and ultra-low friction coatings.

**Acknowledgment**

A.M., G.P., A.S. and A.V. acknowledge support by the Italian Ministry of University and Research through PRIN UTFROM N. 20178PZCB5. A.V. and A.S. acknowledge support by the European Union's Horizon 2020 research and innovation program under grant agreement No. 899285. E.T. acknowledges support by ERC ULTRADISS Contract No. 834402.

**Author Contributions**

GP and AM conceptualized the experimental work and performed the experiments. AS and AK performed the simulations. AS, AV and ET conceptualized the simulations. All authors analyzed the data and wrote the manuscript.

# Supporting information for

# Anisotropic Rheology and Friction of Suspended

# Graphene

*Andrea Mescola[1], Andrea Silva[2,3], Ali Khosravi[2,3,4], Andrea Vanossi[2,3], Erio Tosatti[2,3,4], Sergio Valeri[1,5] and Guido Paolicelli[1]*

[1] CNR-Istituto Nanoscienze - Centro S3, Via Campi 213 41125 Modena, Italy.

[2] CNR-IOM, Consiglio Nazionale delle Ricerche - Istituto Officina dei Materiali, c/o SISSA Via Bonomea 265, 34136 Trieste, Italy.

[3] International School for Advanced Studies (SISSA), Via Bonomea 265, 34136 Trieste, Italy.

[4] The Abdus Salam International Centre for Theoretical Physics (ICTP), Strada Costiera 11, 34151 Trieste, Italy

[5] Department of Physics, Informatics and Mathematics, University of Modena and Reggio Emilia, Via Campi 213 41125 Modena, Italy.

1. **Sample preparation and characterization**

   **- CVD Graphene Deposition**

   **- Sample Cleaning**

   **- Raman Analysis**

2. **AFM and FFM measurements**

3. **Sliding regime / Friction profiles and COF-velocity independency**

4. **Elastic deformation of the membrane**

5. **MD Simulations**

   **- Method**

   **- Finite Size Effect**

   **- Force Friction Protocol**

   **- Force Friction Armchair OBC**

   **- Force Friction Zig-Zag (PBC and OBC)**

# 1. Sample preparation and characterization

**CVD Graphene Deposition:** Commercial single-layer CVD graphene (ACS Materials, Pasadena, CA, USA) was deposited over a SiO2-based calibration grating (TGZ3, NT-MDT Spectrum Instruments) whose pattern's design consists of long parallel grooves spaced by 3 ± 0.05 µm and a step height of 520 ± 20 nm through the standard polymer-assisted wet transfer method. Briefly, by this procedure graphene sheet is provided with one side covered by PMMA polymer and it is supported on a water soluble polymer thick base which makes it easy to handle. Graphene was dipped into DI water taking care of avoiding its rolling up, then left floating at the air-water interface with PMMA side on top for at least 30 minutes to ensure it was fully soaked before transferring it on the substrate of interest. The commercial calibration grating TGZ3 was then used to pick up the floating graphene sheet. Once transferred, the sample was held vertically for few minutes to let excess water flow out, then dried at room temperature for 30 minutes and finally baked at 100°C for 20 minutes.

**Sample Cleaning:** PMMA residues were removed by dipping TGZ3 with graphene on top firstly in hot acetone bath (50°C for 60 minutes), then in hot isopropyl alcohol bath (50°C for 30 minutes); the samples were then baked at 50°C with the blower running on low speed for 20 minutes. The polycrystalline nature of CVD graphene here used which entails the presence of several line defects such as wrinkles and grain boundaries has to be pointed out, as well as PMMA residues that may significantly influence strain, doping and friction force. Nevertheless, Raman analysis revealed the presence of large enough defect free regions (several square micrometers areas) and we never detect suspended bilayer. Finally, all the areas subject to FFM analysis were initially mechanically cleaned from small physisorbed impurities with a sacrificial cantilever and tip different from that used for friction measurement.

After CVD graphene was deposited on TGZ3 grating and properly cleaned, AFM images (tapping mode) were acquired in air, at ambient conditions. The 3D topographical reconstruction in Figure S1 shows a representative sample region including n.2 grooves along which graphene is totally suspended (right side) and n.1 groove devoid of graphene (left side).

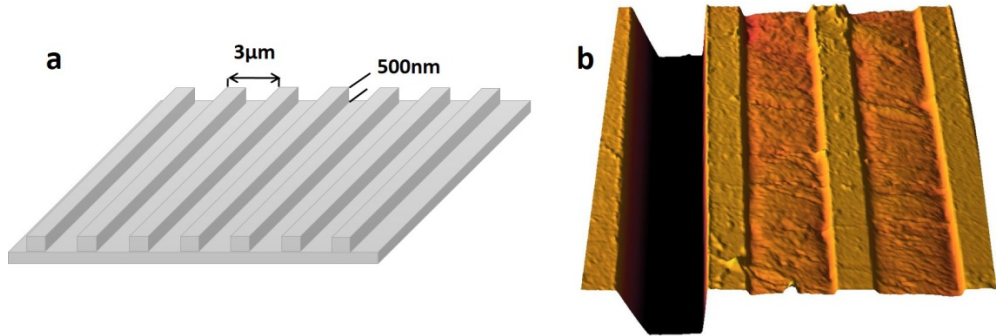

**Figure S 1.** (a) Schematic representation of $SiO_2$ patterned substrate; the pattern's geometry consists of long parallel grooves spaced by 3μm and a step height of 500nm. (b) 3D-AFM image (10x10 μm²).

**Raman Analysis:** Raman spectroscopy measurements were performed using a commercial LabRam micro-Raman spectrometer (Jobin-Yvon, France) equipped with a solid state laser (532nm, maximum power 60mW) and 1800 mm$^{-1}$ grating. All the measurements were performed at 10% laser power (attenuated through ND filters) with 1s exposure at 100X magnification. The acquisition range explored was between 1100 and 3400 cm$^{-1}$.

## 2. AFM and FFM measurements

Topographical analysis as well as friction measurements were carried out by using a commercially available Atomic Force Microscope (AFM) NT-MDT NTEGRA AURA system. All the measurements were performed in air, under ambient conditions (T≈25°C, RH≈55%) using commercially available rectangular shaped silicon cantilevers with nominal elastic constants between 0.3 and 0.8 N m$^{-1}$ (MikroMasch HQ:CSC37/NoAl). Calibration of normal and torsional spring constants were done according to Sader method [42,43]. Friction force loops were typically used to build friction force *vs* normal load plots. In particular, the areas of interest were scanned in two different directions, namely orthogonal and parallel to the grooves axis by rotating the sample of 90° thus keeping the fast scan direction along the x axis. The normal load applied was calibrated by previously acquiring force-distance curves on a flat supported region to calculate the sensitivity factor of the cantilever deflection. The scans were acquired in "one line" mode by decreasing the normal load every forty second (i.e. forty lines at 1Hz scan rate) from ≈25nN to the pull-off value; in this mode the tip continuously slides along the same region net of negligible thermal drift. The friction forces resulting from the difference between forward and backward lateral force signals are then averaged on approximately thirty lines having a constant normal load (far from the set point changes) to produce one data point. All the areas analyzed have been previously mechanically cleaned with a sacrificial cantilever, different from that used for friction measurement, using a normal load (~40nN) greater than the highest load reported in the plot. AFM image processing, including the three-dimensional display of data, was carried out using both the software provided by NT-MDT and the free modular software Gwyddion (version 2.55).

Tip apexes were estimated regularly during FFM measurements by performing a deconvolution of the topography obtained on a AFM calibration sample (PA01/NM) and also post –experiment by scanning electron microscopy (SEM). Typical apex radius is about 15 nm.

### 3. Sliding regime / Friction profiles and COF-velocity independency

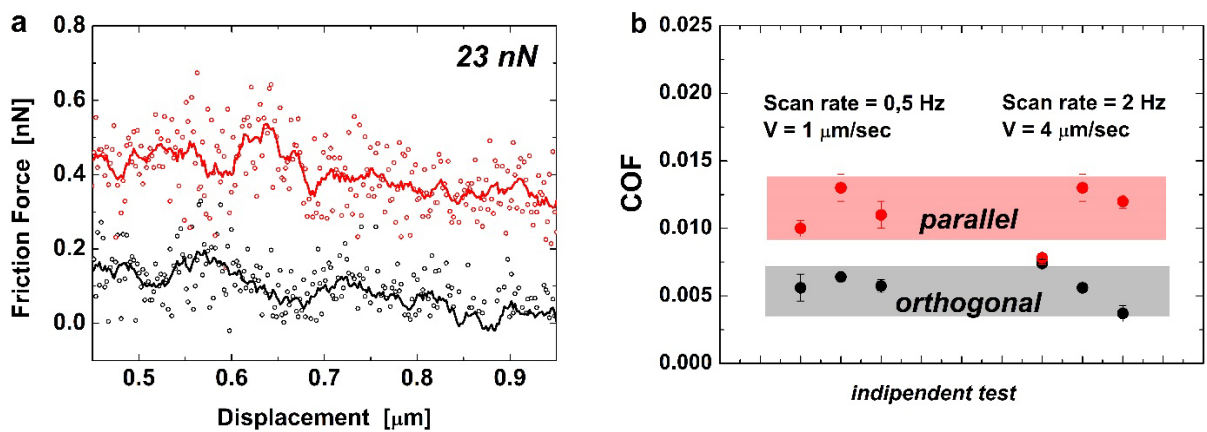

**Figure S 2.** (a)Friction force profiles resulting from parallel (red) and orthogonal (black) scan at constant load of 23nN. Circular dots represent raw data; continuous lines are the relative smoothed signals. (b) COF from parallel (red) and orthogonal (black) scan acquired at different scan speeds, 1 and 4 μm·s⁻¹.

## 4. Elastic deformation of the membrane

Friction vs load curves were evaluated by working in contact mode using single line acquisition and applying a constant normal force. The tip always scanned the same region in back and forth, unless the unavoidable thermal drift, and the applied load was modified by varying the setpoint at regular time intervals maintaining a continuous contact on the surface.

The setpoint is converted to load using the independent measurement of the normal elastic constant $K_N$ obtained previously from the dynamic characteristics of the cantilever and according to the sensitivity factor measured from the force vs distance calibration curve on a rigid portion of the sample. Sensitivity is checked regularly during friction vs load analysis.

To understand the procedure we have used to evaluate the elastic deformation of the graphene membrane, let's consider first a rigid surface. The height profiles on a rigid and flat substrate at different loads correspond to different piezo extensions as shown by continuous lines on figure S3, left. By plotting the applied load as a function of the extension of the piezo for any point of the profile, we obtain the load vs piezo extension curve, right panel of figure S3 (colored dots). As expected, these points completely overlap on the linear region of the load vs distance standard calibration curve (red continuous line) because here piezo elongation just deforms the AFM cantilever.

The same type of analysis gives a different result and it becomes very useful when the surface could deform under load and there are no fixed points along the profile to properly evaluate the induced deformation. This is exactly the case of our FFM measurements on suspended graphene where we measured just at the membrane center without touching the rigid clamping regions.

Figure S4 shows the height profiles of the suspended membrane measured during representative orthogonal (a) and parallel (c) scans in contact mode as a function of applied loads and without ever losing contact. These profiles come simultaneously during friction vs load acquisition (or in general any FFM maps).

In panel b, d we show the load vs distance curve reconstruction by plotting for a particular position of the profile (x= 500 nm) the applied load as a function of the piezo extension (colored dots). In this case the linear portion of the reconstructed curve does not overlap on the corresponding load vs distance calibration curve (red line) and this is a clear indication of substrate deformation.

Fitting the linear region of the reconstructed curve (dash black line) allow us to extrapolate the effective elastic constant ($K_{eff}$) for the system composed of cantilever and graphene which can be thought as two springs in series: being known the elastic constant of the cantilever ($K_N$) it is possible to easily identify that of graphene under strain ($K_{Gr}$). By repeating this argument for all the profile positions (512 pt) we reconstructed the position sensitive $K_{Gr}$ presented on Fig. 6 on the main text.

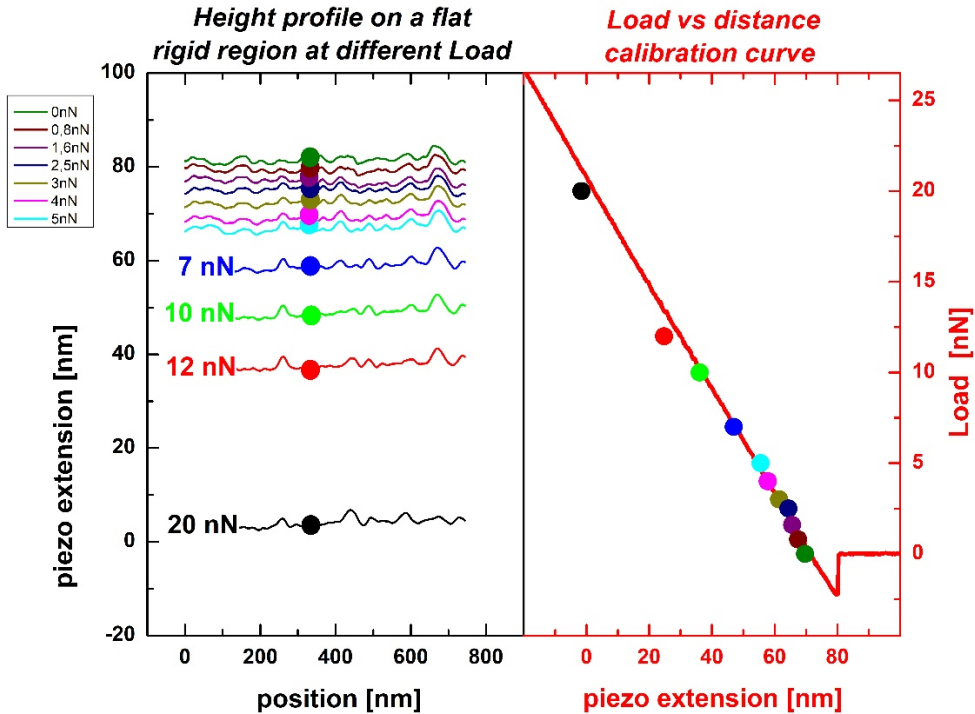

**Figure S 3**. Left, Heights profiles - raw data - as a function of piezo extension at different loads on a flat and rigid substrate. Right, Load vs distance calibration curve (red line) and load vs distance curve reconstruction by plotting for a particular position of the profile (colored dots) the applied load as a function of the piezo extension.

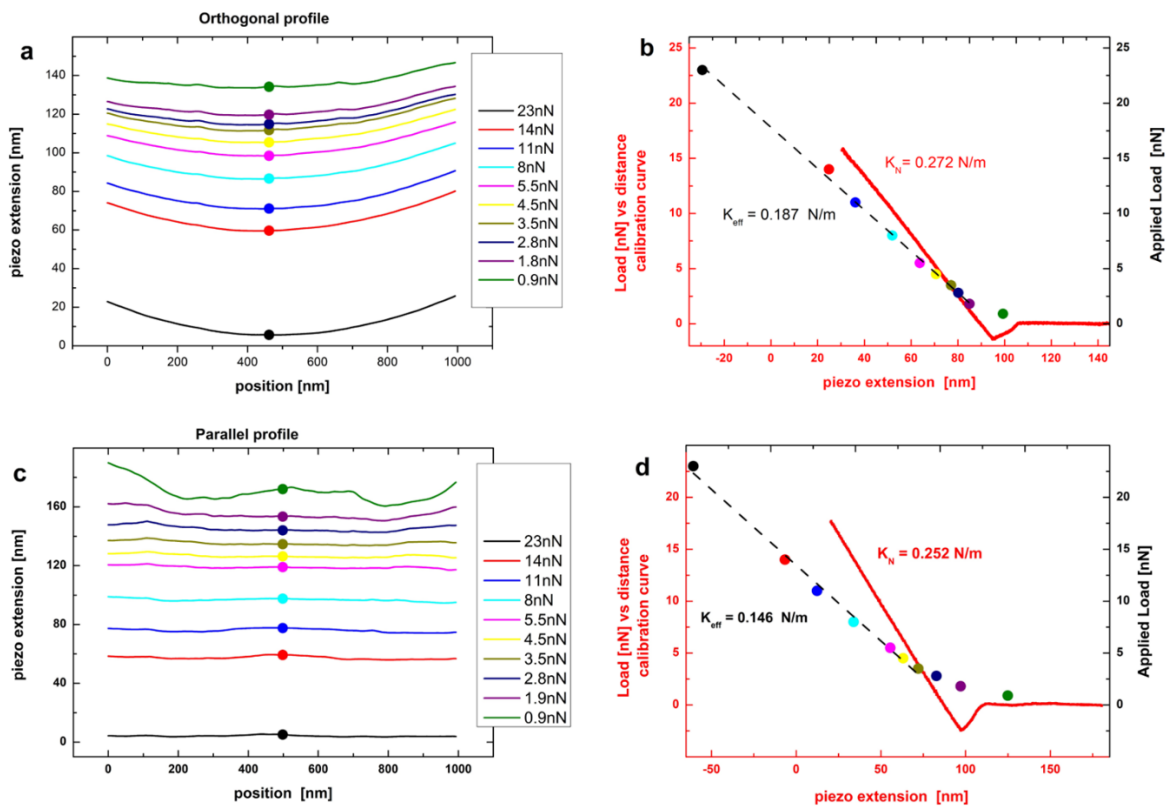

**Figure S 4.** (a) Heights profiles - raw data - collected during a typical FFM orthogonal scans at different loads and presented as a function of piezo extension. (c) Heights profiles - raw data - collected during a typical FFM parallel scans at different loads and presented as a function of piezo extension. (b, d) Load vs distance calibration curve (red line) and load vs distance curve reconstruction by plotting for a particular position of the profile (x= 500 nm, colored dots) the applied load as a function of the piezo extension. The linear regions in panel (b, d) represent the elastic behavior. The slope extracted from calibration curves (red) equals the flexural elastic constant of the cantilever while from the reconstructed curves we evaluated an effective elastic constant of the system. Here we show values extracted at the center of a membrane.

## 5. MD Simulations

**Method.** Non-equilibrium molecular dynamics (NEMD) simulations are performed using LAMMPS [54]. The graphene membrane is modeled using the AIREBO potential [55] and LJ for graphene-Si interaction ε=0.0087 eV and σ=3.595Å [27].

To mimic the suspended graphene, we clamp and hold both sides of graphene in the X-direction with strong springs of stiffness $K_c$ = 1602.17 N/m. This large value is chosen as a compromise between vertical load and size of the system, which is also close to Young's modulus of graphene given its nanometric size. However, the overall features of simulations remain independent of this choice; smaller values would intuitively reproduce the same results for a smaller scale of vertical loads. Simulations with both open and periodic boundary conditions (OBC and PBC) along Y-direction are performed. To test the effect of the membrane crystalline orientation, all measurements are done for both armchair and zigzag orientations of graphene.

The relaxed indentation configurations are obtained in three steps; first, imposing an initial constant vertical load on the tip which allows preliminary shaping during N=3000000 steps of MD simulation with the time step dt=1 fs. Second, using CG minimisation with tolerance $10^{-16}$ eV on the energy allows fast readjustments of the configurations. Third, a final MD run with N=3000000 steps gives us realistic results, allowing for fine readjustment.

To simulate frictional measurements, the system is kept at zero temperature using a Langevin thermostat with damping factor gamma=0.05 ps. We performed more simulations for various values above and below this value, and the results of this work are confirmed to be independent of a specific choice.

To measure friction, the tip is attached to a moving stage with constant velocity $v_{drag}$=2.0 m/s with a spring of stiffness $K_{drag}$~0.08 N/m. The friction force is computed as $F$=$K_{drag}$ $[v_{drag}$ $t$ - $r_i(t)]$ where $r_i(t)$ is the instantaneous position of the tip along the direction $i$, which could be perpendicular ($x$ axis) or parallel ($y$ axis) to the clamps.

Finally, to confirm the consistency of results, the procedure is repeated back and forth many times to regain the same hysteresis.To estimate the contact area of atomically-thin tip adopted in MD simulation of the PT model, we adopt a simple Hertian model. The radius $a$ of the contact area is expressed in terms of the applied load $P$ and mechanical properties of the tip by $a = (\frac{3 P R}{4 E^*})$ , where R is the curvature radius of the tip, here the LJ interaction R=0.35 nm, and the effective Young modulus of the contact is given by $1/E^* = (1 - \nu_1^2)/E_1 + (1 - \nu_2^2)/E_2$. The Young and Poisson moduli for graphene and silicon tip, respectively, are taken from the Material Project [56,57] : $E_1 = 371$ GPa, $\nu_1 = 0.19$ $E_2$ =151 GPa, $\nu_2$=0.20.

**Finite size effect.** As we perform zero temperature friction simulation, the tip might get trapped in unstable configurations along high-symmetry directions. In order to break the symmetry along these directions, a small temperature of 0.001 K is added to the tip only: this allows the tip to jump the center of the graphene hexagons when dragged along the armchair direction ($x$ axis), resulting in the well-document stick-slip zig-zag trajectory.

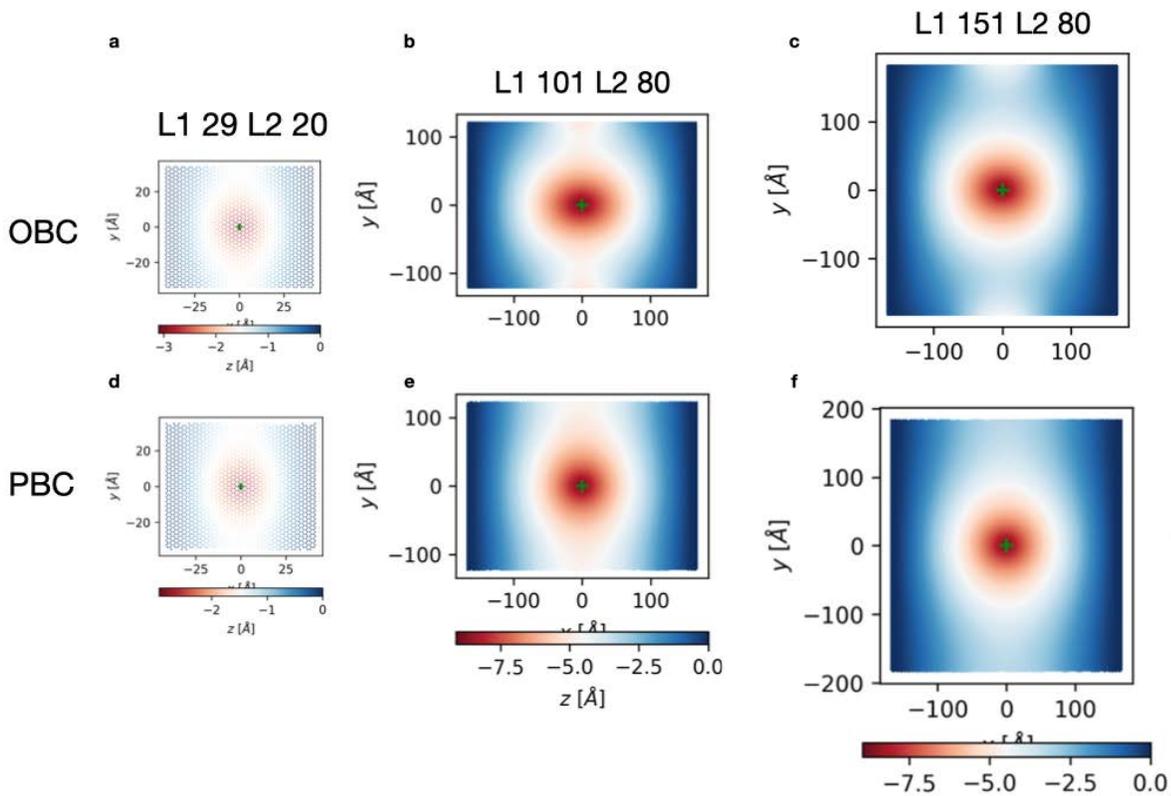

**Figure S 5.** Size scaling of the indentation pattern in MD simulations. (a,b,c) refer to open boundary conditions (OBC), (d,e,f) to periodic boundary conditions (PBC). The size of each system is reported on top of each plot in repeated graphene units along x and y, respectively. The color reports out-of-plane deformation; the scale is reported in the colorbars below each plot.

**Force Friction Protocol.** When sliding in the perpendicular direction, the membrane energy presents an additional potential energy contribution due to the elastic deformation induced by the clamps. When going from the center of the membrane to the side, this potential energy increases; upon reversing the force direction, the elastic deformation is reversed and this potential energy is recovered, see Fig S6 a below. This potential energy results in a linear shift of the force recorded by the AFM, see Fig S6 b. The dissipated energy is the area enclosed within the force loop in Fig S6 b, which is independent of this potential energy term: this is clear when comparing the area enclosed by the raw force signal (dotted curves) and the shifted force one (solid curves) in Fig S6 b. To simplify and speed up the computational protocol, the force is recorded only in the forward direction and then shifted according to the linear envelope of the force, as shown in Fig S6 b. In this way the area of the friction loop is preserved but the average friction force of the shifted curve is proportional to the dissipation

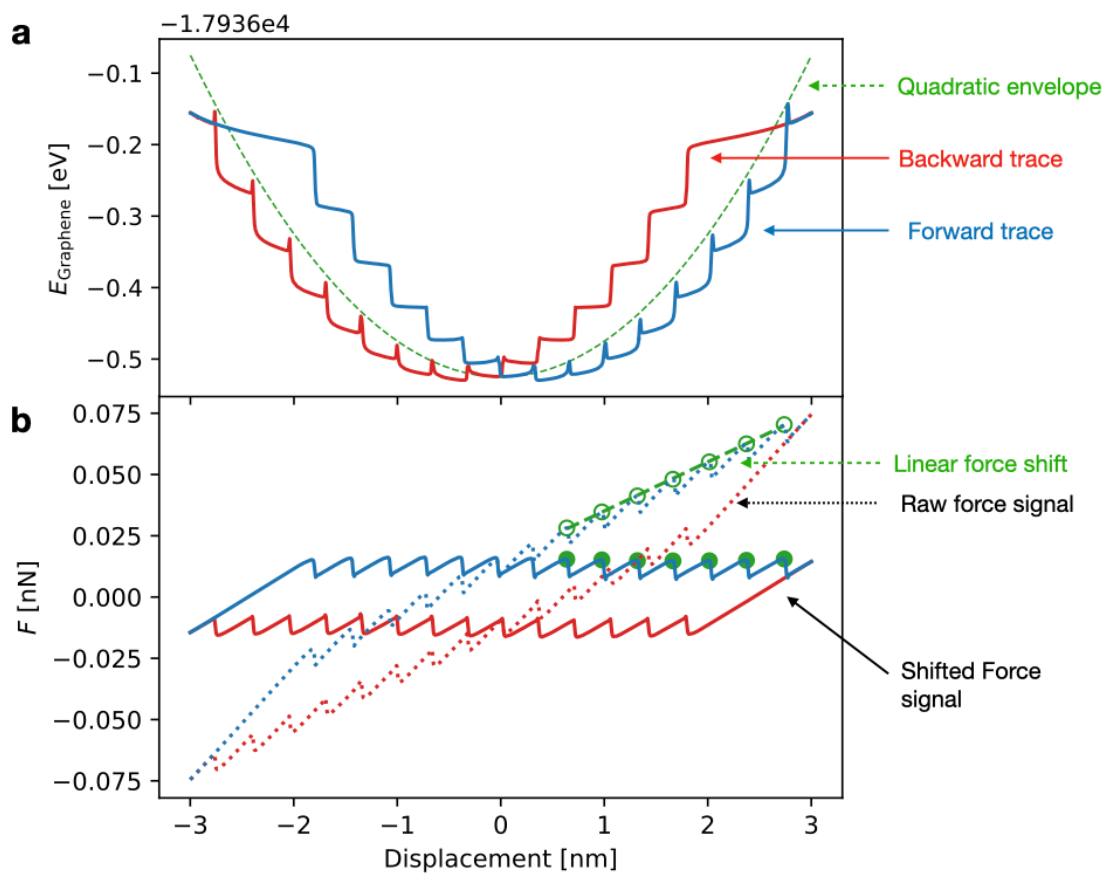

**Figure S 6.** Example of the protocol adopted to compute the friction traces from MD simulations. (a) The ordinate reports the energy of the system and the abscissa the displacement during the forward (blue line) and backward (red line) traces. The green dashed line shows a quadratic function as a guide to the eye. (b) Force traces relative to the same forward and backward traces. The dashed lines report the raw, unshifted traces. The green dashed line shows the linear envelope of the peaks of the raw forward trace (empty circles). The solid lines report the shifted traces.

**Armchair OBC.** Simulations with OBC are in agreement with the PBC protocol, as shown in Fig. S7. The coefficient of friction extracted from the force-load curve are COF$_x$=0.008 COF$_y$=0.012, resulting in a ratio COF$_y$/COF$_x$ =1.615

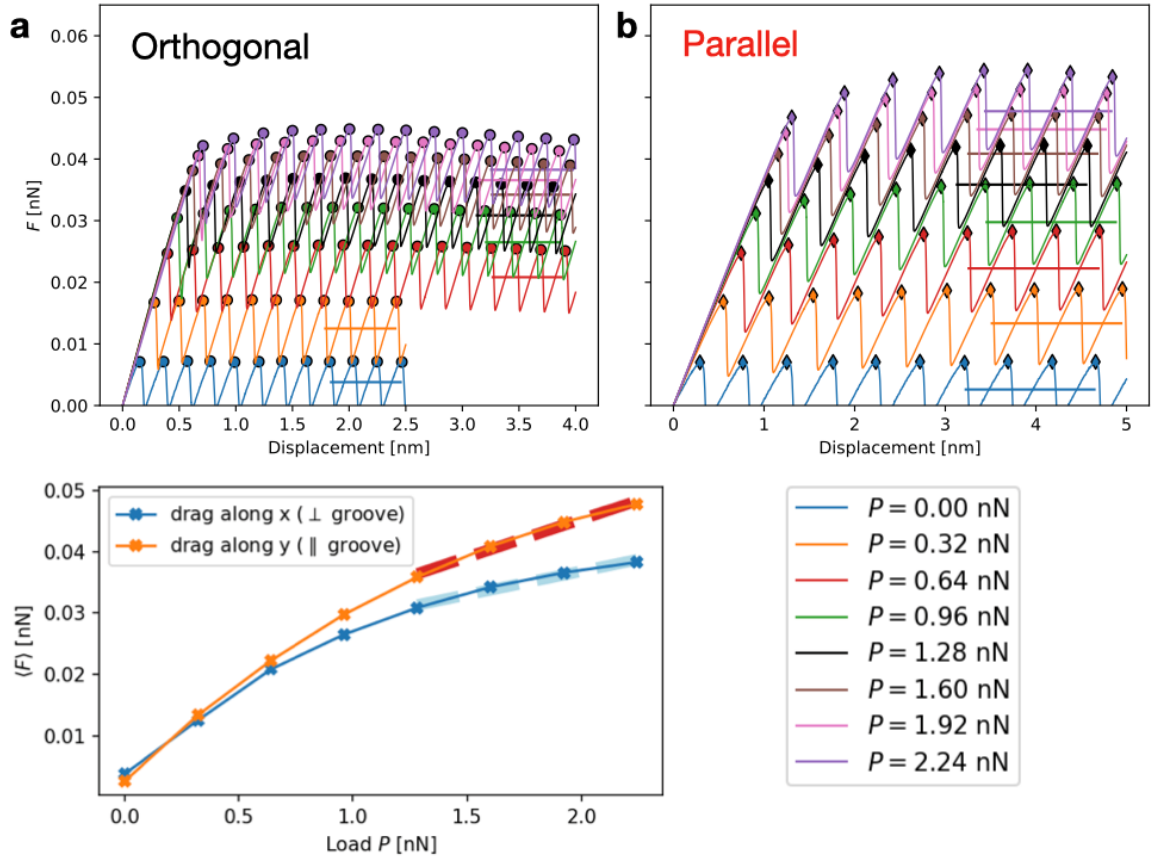

**Figure S 7.** Force traces (a,b) and friction-versus-load (c) for the armchair OBC system. The figures follow the same convention as Figure 4 and Figure 5a in the main text.

**Zig-Zag system (PBC and OBC)**

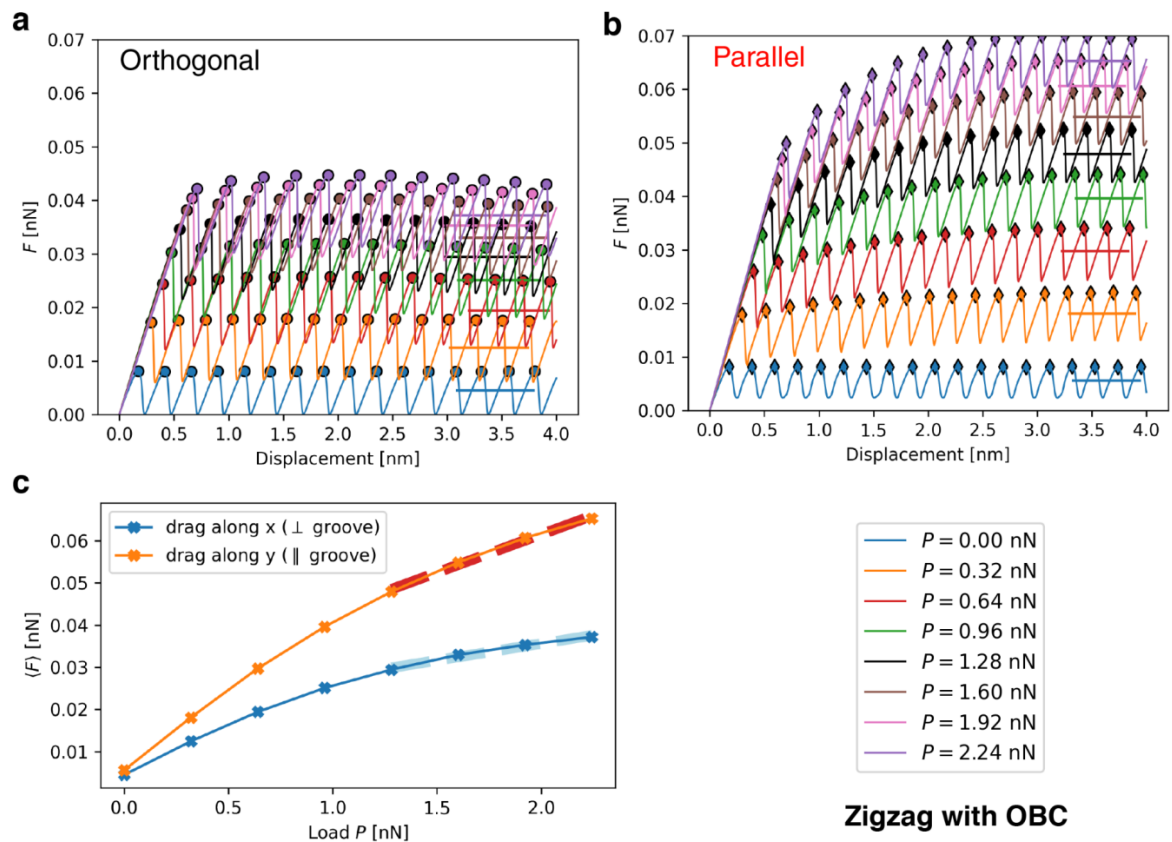

**Figure S 8.** Force traces (a,b) and friction-versus-load (c) for the zig-zag OBC system. The figures follow the same convention as Figure 4 and Figure 5a in the main text.

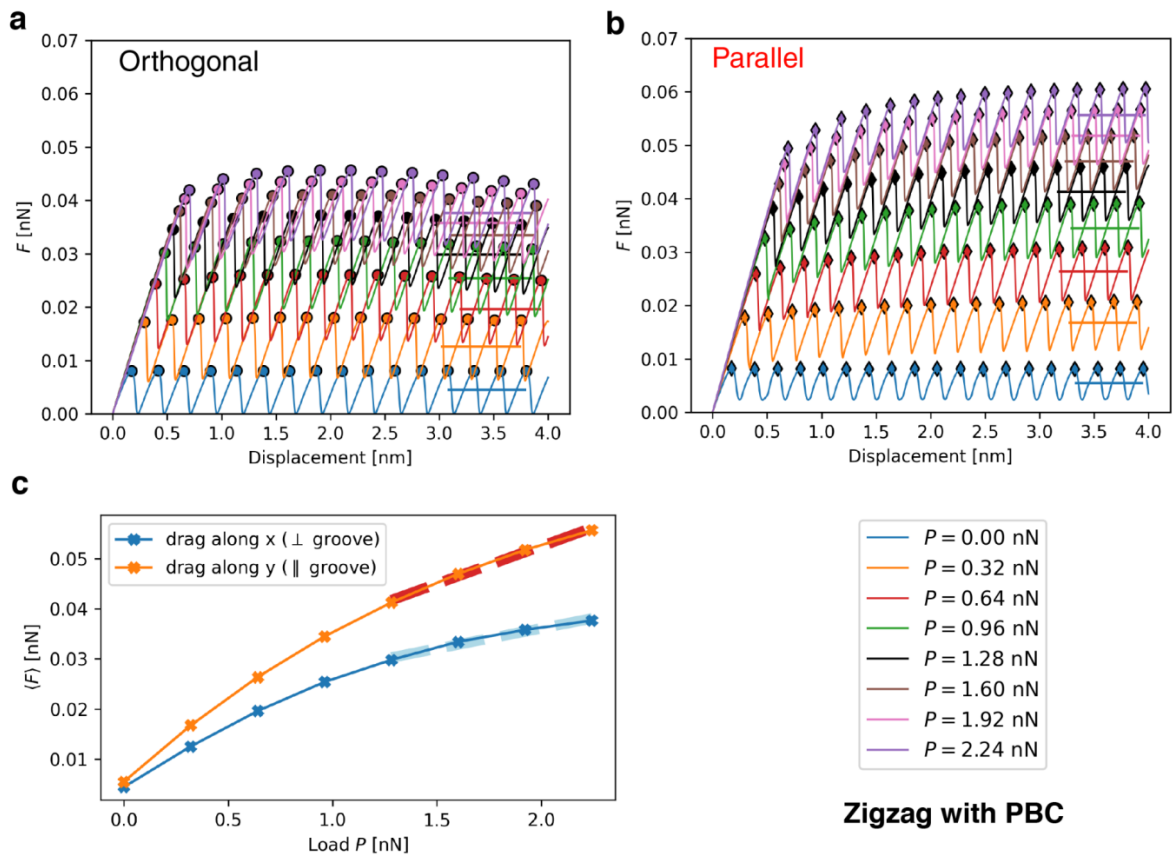

**Figure S 9.** Force traces (a,b) and friction-versus-load (c) for the zig-zag PBC system. The figures follow the same convention as Figure 4 and Figure 5a in the main text.

From simulations we obtain the ratio COFy/COFx for zigzag orientation with OBC and PBC, 2.273 and 1.851 respectively. These results confirm higher friction along Y-direction, as observed previously for armchair orientation; 1.615, and 1.763 –for OBC and PBC. In agreement with known dissipation asymmetry due to crystallographic orientation[59]. Moreover, the largest ratio of friction coefficients for zigzag with OBC could be ascribed to the, which remain valid, simply because zigzag boundary having ahas larger bond density than armchair (0.85 C-C and 0.75 C-C): this, making the boundary looser and, resulting in higher friction.

| Ratio COFy/COFx | Armchair | Zig-zag |
|---|---|---|
| OBC | 1.6 | 2.3 |
| PBC | 1.7 | 1.9 |

**Table S1.** Ratio COFy/COFx for all the four systems in MD.